\documentclass[aps,twocolumn,groupedaddress,prb,floatfix,showpacs]{revtex4}
\usepackage{graphicx}
\usepackage{dcolumn}
\usepackage{bm}
\usepackage{amsmath}
\usepackage{verbatim}

\begin{document}

\title{Density Functional Study of Ternary Topological Insulator Thin Films} 

\author{Jiwon Chang}
\email{jiwon.chang@mail.utexas.edu}
\author{Leonard F. Register}
\author{Sanjay K. Banerjee}
\author{Bhagawan Sahu}
\affiliation{
Microelectronics Research Center, The University of Texas at Austin, Austin Texas 78758
}

\date{\today}

\begin{abstract}
Using an {\it ab initio} density functional theory based electronic structure method with a semi-local density approximation, we study thin-film electronic properties of two topological insulators based on ternary compounds of Tl (Thallium) and Bi (Bismuth). We consider TlBiX$_2$ (X=Se, Te) and Bi$_2$X$_2$Y (X,Y= Se,Te) compounds which provide better Dirac cones, compared to the model binary compounds Bi$_2$X$_3$ (X=Se, Te). With this property in  combination  with  a  structurally  perfect  bulk  crystal, the latter ternary compound has been  found  to have  improved surface  electronic  transport in recent  experiments. In this article, we discuss the nature of surface states, their locations in the Brillouin  zone and  their interactions within the bulk  region. Our calculations suggest a critical thin film thickness to maintain the Dirac cone which is significantly smaller than that in binary Bi-based compounds. Atomic relaxations or rearrangements are found to affect the Dirac cone in some of these compounds. And with the help of layer-projected surface charge densities, we discuss the penetration depth of the surface states into the bulk region. The electronic spectrum of these ternary compounds agrees  very well with the available experimental results.    
\end{abstract}

\maketitle

\section{Introduction}
Three  dimensional  (3D)  Topological  band  Insulators (TI)  have attracted considerable  attention from the con- densed  matter and  device  physics  communities for the novel electronic surface states they support and the host of unusual responses to external fields\cite{zahid}. The basic understanding of their electronic structure, using Density Functional Theory (DFT) is a continuing quest in the literature. These studies have contributed to the growth of the field and  helped  with interpreting experiments.   In recent  years,  binary  3D TI  materials  Bi$_2$Se$_3$ and Bi$_2$Te$_3$ have  emerged  as model systems for numerous experiments which  focused  on exfoliation and  the molecular  beam  epitaxy growth of thin films and for probing the electronic structure by optical techniques\cite{zahid2} and device fabrications\cite{mit}. The theoretical works, based on density functional theory or tight-binding calculations, focused on elucidating the structure-property relations, predicting new TIs, and mapping the band structures\cite{louie}. One of the concerns for the use of binary TI materials in devices is the intrinsic {\it n}(or {\it p})-type vacancies in the bulk crystals of Bi$_2$Se$_3$ (Bi$_2$Te$_3$) which make the conduction through bulk states dominate transport experiments\cite{ong} and the hexagonal {\it warping} effect that makes the Dirac cone anisotropic, especially in the conduction band region. This anisotropic nature of the Dirac cone leads to interesting spin-textures (or spin-momentum locking) of surface states with possible applications in spintronics and quantum information processing\cite{bansil}. Recently, two promising 3D TI materials, TlBiX$_2$ (X= Se,Te) and Bi$_2$X$_2$Y (X,Y = Se,Te) have been predicted to have near perfect Dirac cones, in terms of less entanglement of bulk and surface states, and experiments have supported these predictions.\cite{ando,lin,souma,xu,chen}. Moreover, Bi-based ternary compounds offer high bulk resistivity, due to the structurally perfect nature of the crystals, so surface transport is enhanced. 

\begin{figure}[ht!]
\scalebox{0.40}{\includegraphics{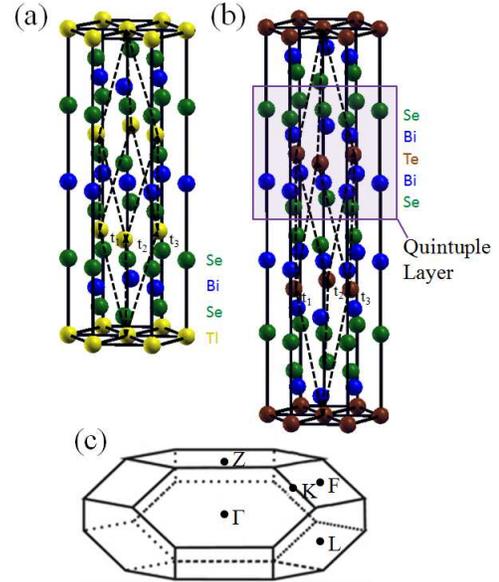}}
\caption{ (Color online) Schematic of the hexagonal bulk structures of representative ternary compounds (a) TlBiSe$_{2}$ and (b) bulk Bi$_{2}$Se$_{2}$Te derived from their corresponding bulk trigonal structures whose three primitive vectors are denoted by t$_1$, t$_2$ and t$_3$. The trigonal structure is part of the larger hexagonal structure shown by dashed lines. In both the compounds, the hexagonal cell contains three times the number of atoms compared to that in the trigonal structure. In (b), the five atomic layers, forming a quintuple layer are shown in the shaded square region. (c) The first Brillouin Zone of the bulk trigonal structure with four time-reversal invariant points $\Gamma$, Z, F and L are shown.}
\label{fig:Fig1}
\end{figure}

In view of these promising advances, a comprehensive theoretical study of the thin-film structures of these ternary 3D TI is necessary to help better understand these materials and to help design future experiments to verify these effects. These materials can have intrinsic size limits which can protect the metallic nature of the surface bands, the surface states can spread inside the bulk region, and atomic rearrangements in thin layers can have profound effect on the Dirac cone itself. We address these issues in this paper using a  DFT-based electronic structure method and compare our results with available experimental results as well as with the results of binary Bi-based TIs. Our studies will have implications for the understanding of topological surface states in ternary TIs and their intended applications. 

Our paper is organized as follows. In Section II, we describe the bulk crystal structures of TlBiX$_2$ (X=Se,Te) and Bi$_2$X$_2$Y (X,Y=Se, Te), their conduction band (CB) and valence band (VB) structures and the computational method used for this study. In Section III we present the thin film electronic structure of these materials and discuss the role of atomic relaxations or rearragements, resulting from thin film formation from the bulk crystal, on the shape and size of the Dirac cone in the bulk, and predict critical film thicknesses required to maintain the Dirac cone  and degree of surface state extension into the bulk region which have yet to be determined experimentally for these class of materials. We compare these results with those available for binary TI compounds. Finally, in Setion IV, we present our summary  and conclusions.

\section{Computational method and bulk band structures}

This section details the computational method used, the choice of computational parameters, the bulk crystal structures and the resulting electronic band structure. The bulk crystal structures of both Tl and Bi-based ternary compounds are similar to the binary compounds Bi$_2$X$_3$(X=Se,Te). Both compounds have a trigonal structure with covalently bonded alternating cations and anion layers stacked along the crystallographic {\it z}-direction. However, there is one important difference as compared to the binary compounds: The atomic layers of these Tl-based structures are not arranged in a quintuple-like (QL) order, suggesting covalent bonding between the unit-cells, and therefore, that exfoliation techniques, used to peel binary TI flakes from the corresponding bulk crystals, cannot be used for extractng Tl-based TI thin films. The layers are arranged in the order Tl-Se(Te)-Bi-Se(Te). The trigonal unit cell contains four atoms (as opposed to five in binary Bi-based TIs) with the lattice parameter {\it a} = 0.7887 nm(0.8263 nm) and the angle between the lattice vectors spanning the lattice of $\alpha$$\sim$31.4$^o$($\sim$31.8$^o$) for TlBiSe$_2$(TlBiTe$_2$)\cite{mahanti}. The hexagonal cell, formed from the trigonal, consists of twelve atomic layers with the number of atoms tripled (Fig. 1(a)). The lattice parameters of the hexagonal cell are  {\it a}=0.4264 nm(0.4534 nm) and {\it c}=2.2478 nm(2.3512 nm) for TlBiSe$_2$(TlBiTe$_2$) and the layer sequence in this case is same as the trigonal cell. Compared to the corresponding binary compound, the Bi-based ternary compound Bi$_2$Se$_2$Te (Bi$_2$Te$_2$Se) has one of the Se (Te) layers replaced by a Te (Se) layer, and has the layers arranged in the order Te(Se)-Se(Te)-Bi-Bi-Se(Te). The QL-like layer arrangement in these compounds facilitates exfoliation-like methods which can be used to peel thin films from the bulk-crystals. The lattice parameters of the bulk trigonal cell is almost same for Bi$_2$Se$_2$Te (Bi$_2$Te$_2$Se): {\it a} = 1.0046 nm(1.0255 nm) and $\alpha$ $\sim$ 24.2$^o$ ($\sim$ 24.1$^o$)\cite{nakajima}. The hexagonal cell, built from the trigonal structure, contains fifteen atomic layers with the lattice parameters {\it a} = 0.422 nm(0.428 nm) and {\it c} = 2.92 nm(2.99 nm) for Bi$_2$Se$_2$Te (Bi$_2$Te$_2$Se) (Fig. 1(b)).     


We used a DFT-based electronic structure method with projector-augmented wave basis\cite{vasp1} and a generalized gradient approximation to the exchange-correlation potential\cite{perdew} for computing the electronic properties of the bulk and thin films of both of the compounds. Spin-orbit coupling (SOC) was invoked in the calculation as implemented in the numerical method\cite{vasp2}. For the bulk structural optimizations of both the compounds, a kinetic energy cut-off of 400 eV and a non-orthogonal {\bf k}-point mesh along the reciprocal lattice vectors of 9 $\times$ 9 $\times$ 9 in the first Brillouin zone (BZ) were chosen. Since previous structural studies of bulk trigonal structures reported agreement with experimental lattice parameters\cite{expt}, we chose to relax only the internal parameters, namely the atomic positions, keeping the lattice constants fixed to experimental values. The total energy is assumed to be converged when all the components of Hellman-Feynman forces on each ion are smaller than the threshold 0.001 eV/\AA. Convergence of computed properties were carefully checked with respect to large energy cut-off, {\bf k}-point mesh and larger force threshold. For both the compounds, the computed interlayer distances are quite close to the experimental works\cite{expt}. These computed distances suggest van der Waals type of bonding between the adjacent QLs in ternary Bi-based TI but not in Tl-based TIs.  With these informations, we built the bulk hexagonal cell and the corresponding thin-films of both the compounds as detailed in the next section.

\begin{figure}[ht!]
\scalebox{0.26}{\includegraphics[angle=-90]{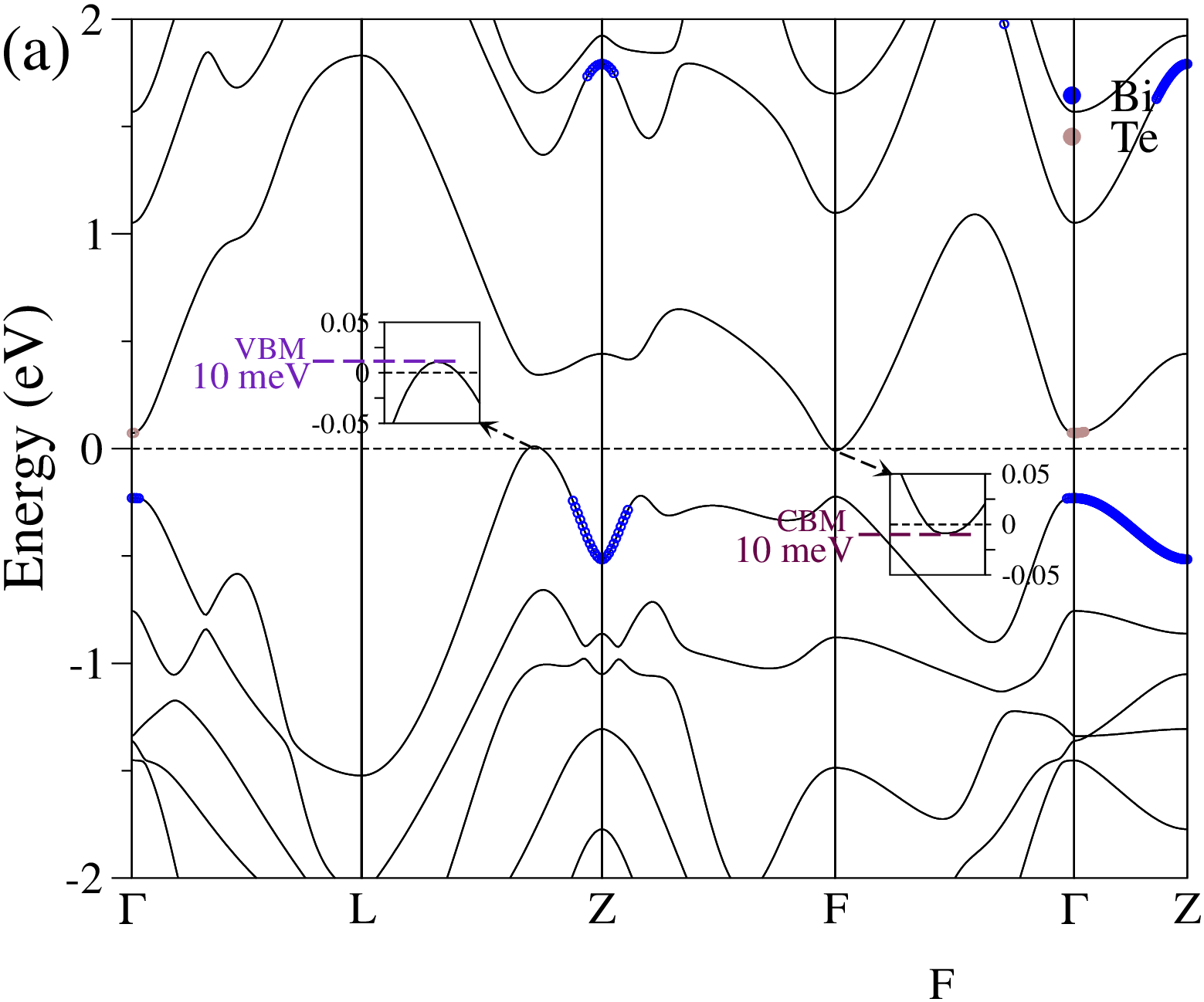}}\\
\scalebox{0.26}{\includegraphics[angle=-90]{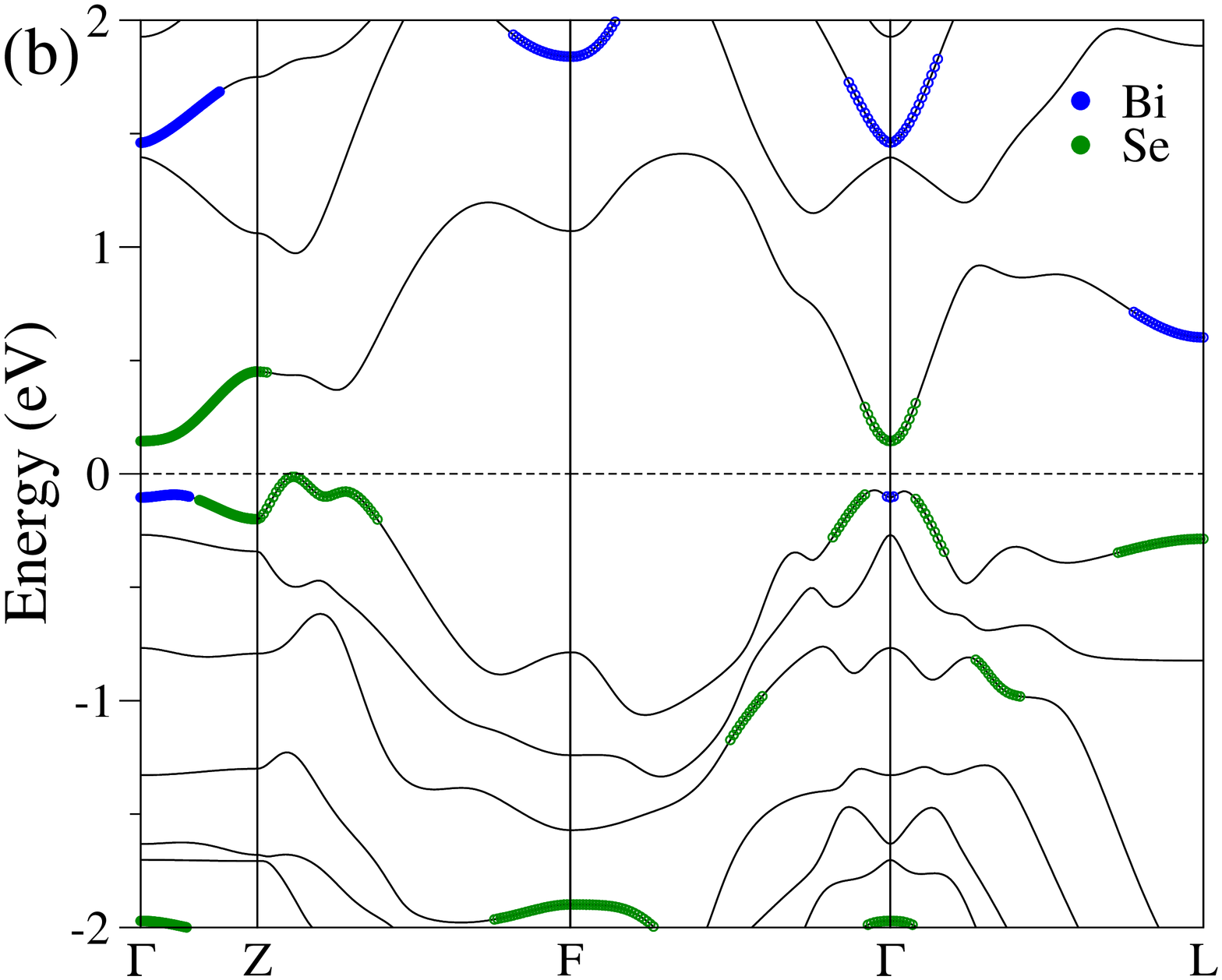}}
\scalebox{0.26}{\includegraphics[angle=-90]{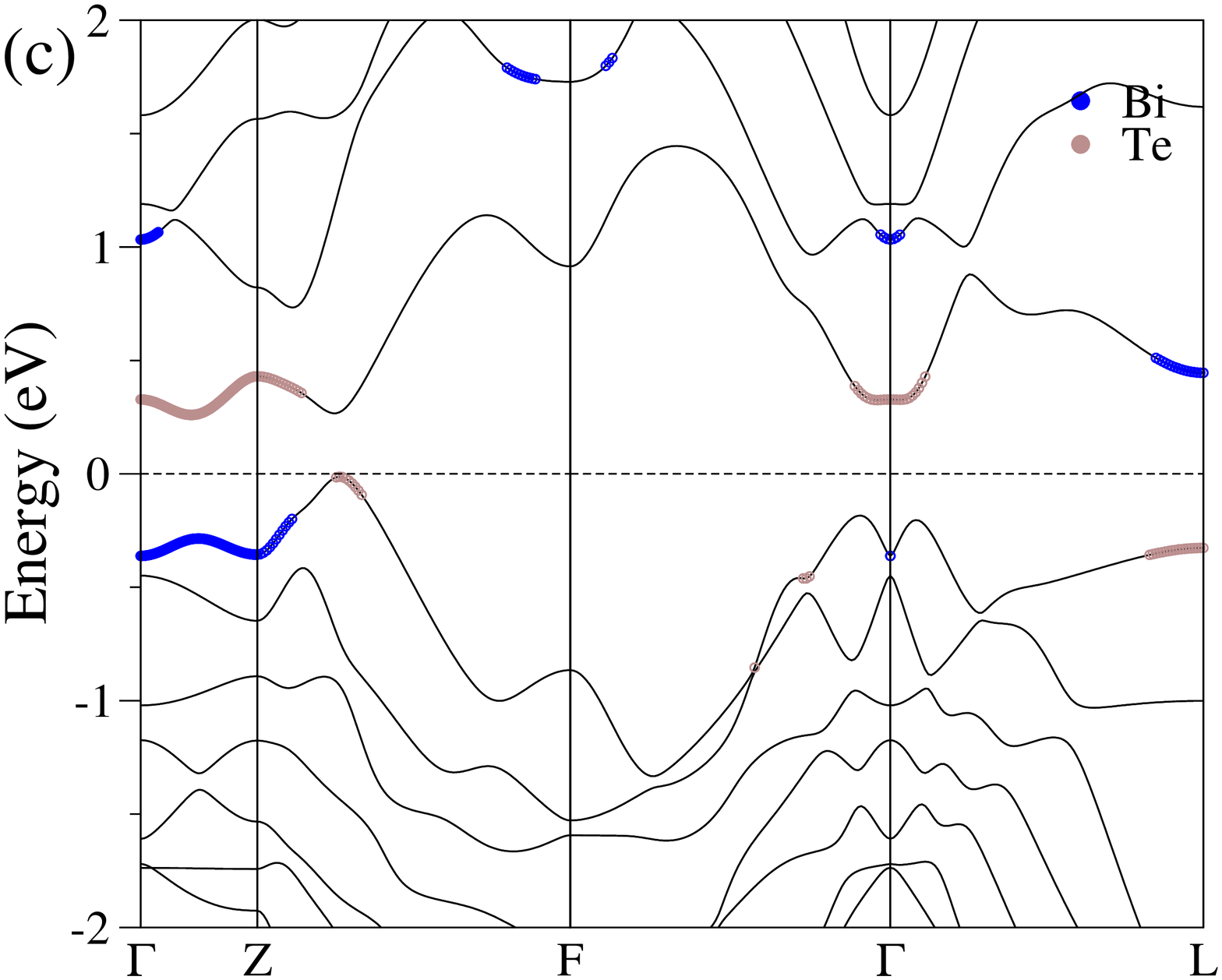}}\\
\caption{ (Color online) Band structure of bulk trigonal structure of (a) TlBiTe$_{2}$ and (b) Bi$_2$Se$_2$Te and (c) Bi$_2$Te$_2$Se with spin-orbit coupling, shown along high symmetry dirctions in the bulk BZ. We obtained a negative indirect gap in TlBiTe$_2$ and indirect gaps in Bi-based ternary TIs. Energy states mainly from $p_z$ orbitals of Bi, Se and Te  are marked on the top of bulk bands and the procedure to obtain these surface bands from crystal wave-functions is discussed in the text.}
\label{fig:Fig2}
\end{figure}

\begin{figure}
\scalebox{0.26}{\includegraphics[angle=-90]{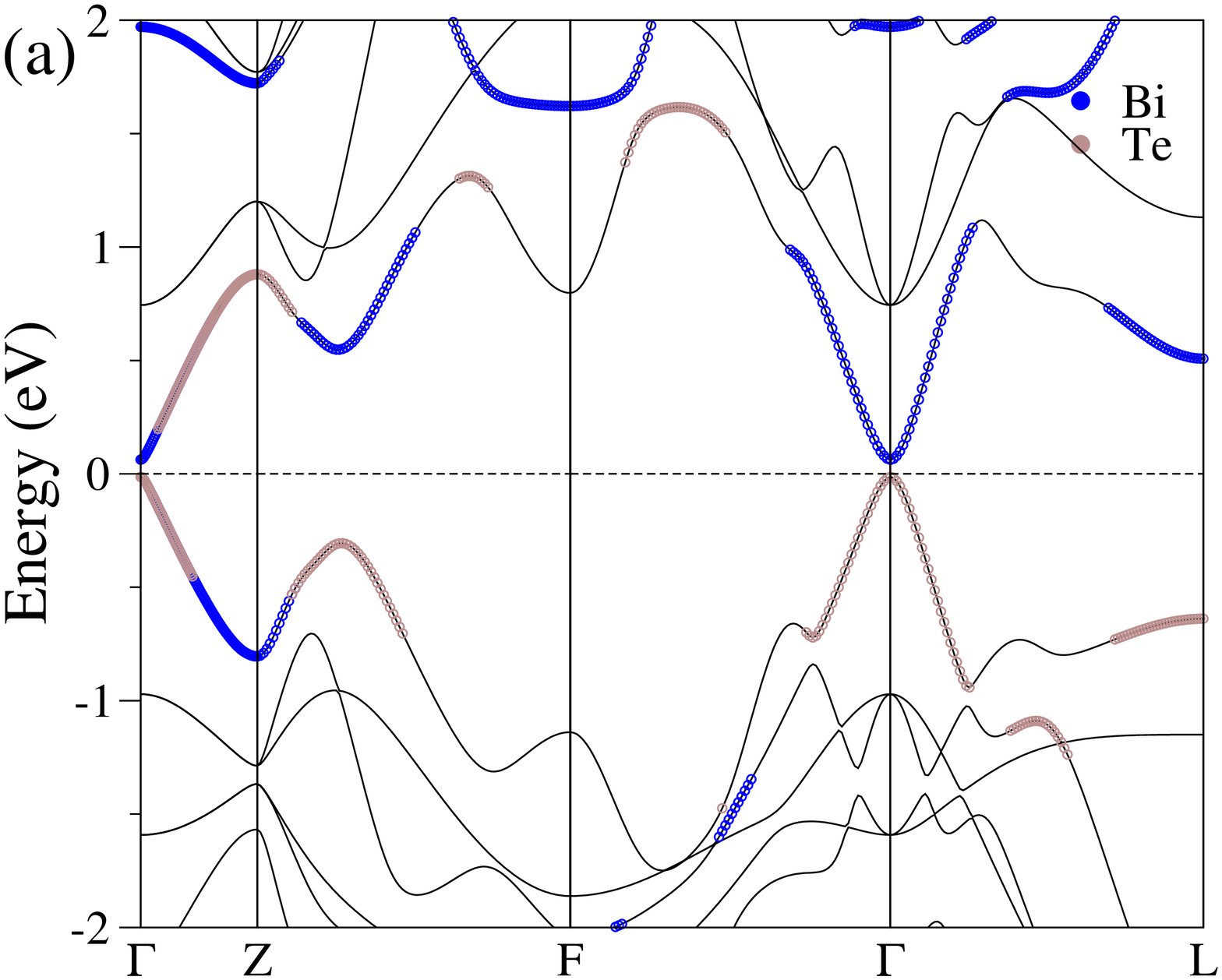}}
\scalebox{0.26}{\includegraphics[angle=-90]{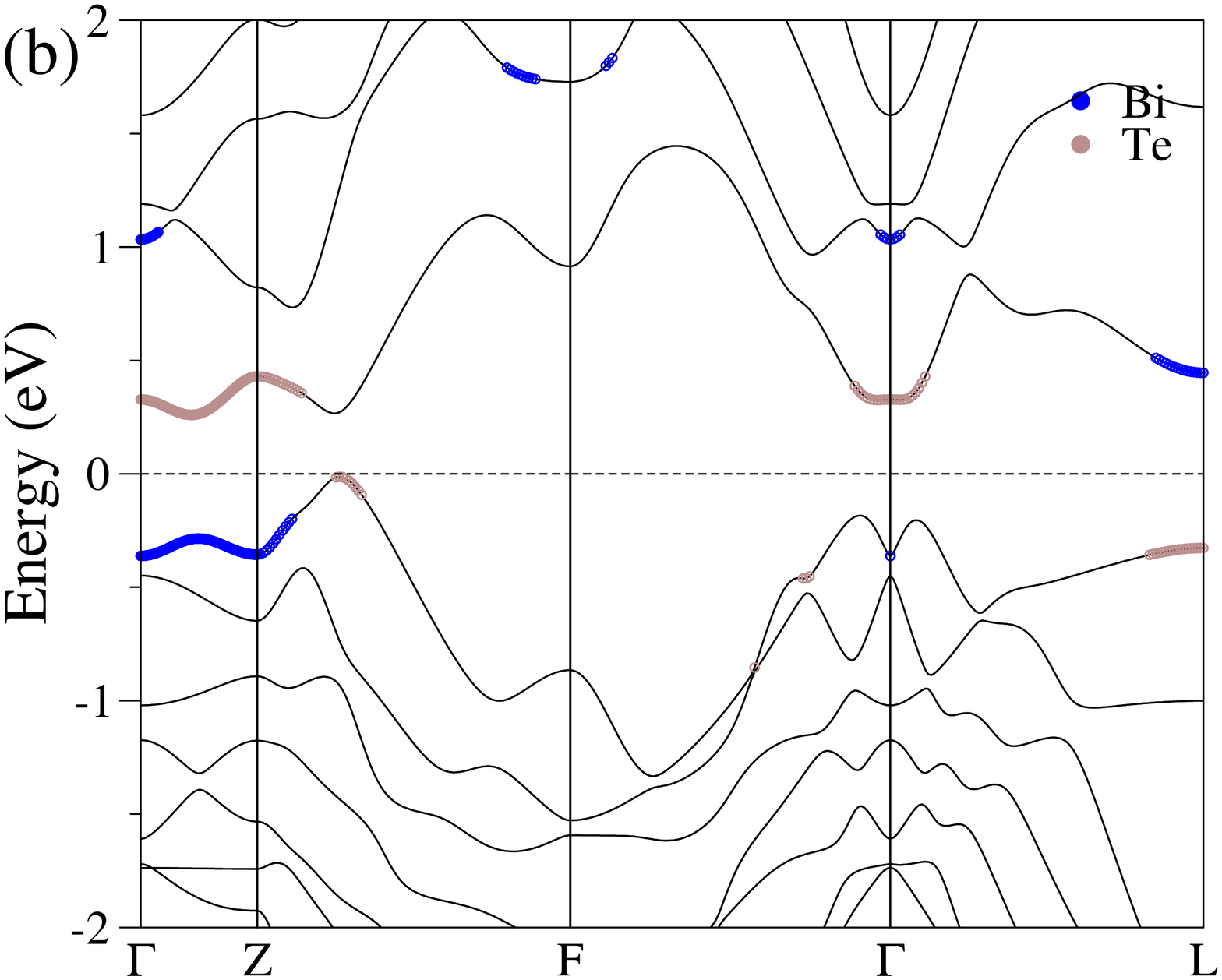}}
\caption{ (Color online) Band structure of bulk Bi$_{2}$Te$_{2}$Se along high symmetry directions (a) without spin-orbit coupling, and (b) with spin-orbit coupling. Energy states formed mainly from $p_z$ orbitals of Bi and Te are marked on the top of bulk bands and band inversion processes are clearly shown.}
\label{fig:Fig3}
\end{figure}

The bulk band structures of TlBiX$_2$ (X= Se,Te), without and with SOC, agree well with other theoretical calculations\cite{mahanti, mahanti2}. However, with SOC, other calculations predict TlBiTe$_2$ 
to be an {\it indirect} semiconductor with the CB minimum (CBM) at the $\Gamma$ point and the VB maximum (VBM) on the line joining the {\bf L} and {\bf Z}-points. The high-symmetry points are labeled in Fig. 1(c). Recent experimental studies argue TlBiTe$_2$ to be a {\bf semimetal} with an negative energy gap of 20 meV\cite{chen}. Our results agree well with this experimental work (Fig. 2(a)). TlBiSe$_2$ has a direct gap at $\Gamma$ of 124 meV(235 meV) without(with) SOC. Figure 2(b) and (c) show the bulk band structures of Bi$_2$X$_2$Y (X,Y=(Te,Se) or,(Se,Te)) with SOC, which agree quite well with a recent DFT calculation\cite{johnson}. The CBM is at the $\Gamma$-point and the VBM is between $\bf Z$ and $\bf F$-points for both the Bi-based TIs. The computed {\it indirect} gap are 157 meV and 272 meV, for Bi$_2$Se$_2$Te and Bi$_2$Te$_2$Se, respectively. We note that the gap is {\it direct} at the $\Gamma$-point without including SOC with the gap values of 633 meV and 76 meV, for Bi$_2$Se$_2$Te and Bi$_2$Te$_2$Se, respectively.

In the literature, emergence of these topological surface states within the bulk band gap of binary 3D TIs was associated with the bulk {\it band inversion} when SOC is switched on\cite{zahid}. We tested this conjecture by plotting the specific orbital contributions from a given atom as a function of {\bf k} and band on the band structure diagrams of Figs. 2 and 3. These specific contributions are normalized with respect to the contributions from all the orbitals in both the bulk Tl- and Bi-based ternary compounds. Since the bulk CBM and VBM at $\Gamma$ consist mainly of $p_z$ orbitals\cite{mahanti2} of the Bi and Se(Te) atoms, respectively for Tl-based compounds, this state was chosen for the orbital projection studies. In Tl-based TIs, a cut-off contribution percentage is chosen which gives a reasonable picture of {\it band inversion} in these figures. If the calculated $p_z$ orbital contribution of a specific atom at a certain state is greater than a given cut-off percentage, we considered this state as mainly originating from that atom. This cut-off was set at 50$\%$ and 25$\%$ for Se(Te) and Bi atoms, respectively. A slight change in a cut-off percentage, below or above these choices, resulted in an insignificant change for the CBM and VBM bands at the $\Gamma$-point, which is region of interest to study the {\it band inversion} process. For example, the $p_z$ orbital contributions on the band structure of TlBiTe$_2$ in Fig. 2(a) remain similar near the CBM and VBM around the $\Gamma$-point with the cut-off percentage of 45$\%$ and 22.5$\%$ for Te and Bi atoms, respectively (Figures not shown). A similar {\it band inversion} effect was also seen in Bi-based TIs (Bi$_2$Te$_2$Se and Bi$_2$Se$_2$Te). This effect is shown only for one of the compound Bi$_2$Te$_2$Se (Fig. 3(a) and (b)). The other compounds show similar effects (Figures not shown). We gave an equal weight of 30$\%$ to both Bi and Te(Se) orbital contributions since the two outermost layers of Se(Te) and Bi, out of the total of five atomic layers within each QL, mainly particpate in forming the VBM and CBM.      

\begin{figure}
\scalebox{0.550}{\includegraphics{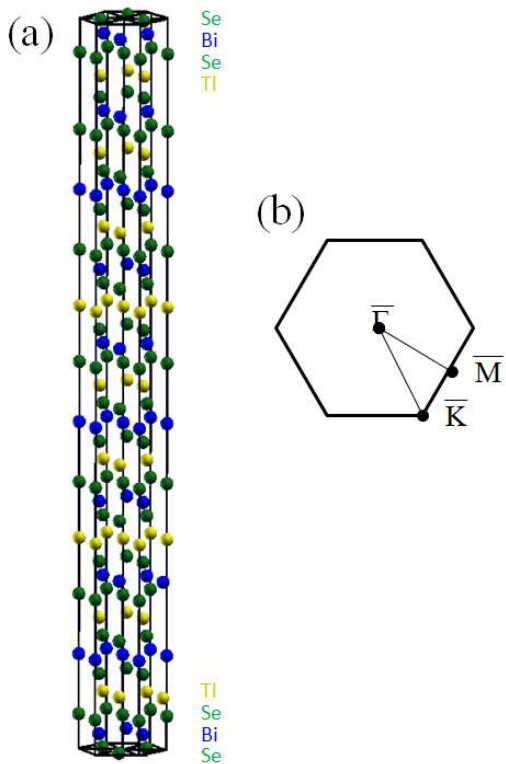}}
\caption{ (Color online) Schematic diagram of thin film structures of (a) TlBiSe$_{2}$ obtained by stacking 39 layers along the {\it z}-direction. The atomic layers close to surfaces are in the order of Se-Bi-Se-Tl, on both sides of the film. (b) Two-dimensional Brillouin zone of the (111) surface of the bulk TlBiSe$_{2}$ with three time-reversal invariant points  $\bar{\Gamma}$, \={M}, and \={K}
}
\label{fig:Fig4}
\end{figure}

\section{Thin film band structures and surface states}

In this section we discuss the surface states of ternary TIs. First we discuss the interesting nature of surface states in Tl- and Bi-based TIs. We compare our results with binary Bi-based TIs, wherever possible. We choose kinetic energy cut-off of 400 eV and {\bf k}-mesh size of 9 $\times$ 9 $\times$ 1 on the surface BZ of the hexagonal supercell for computing surface band structure of both Tl- and Bi-based ternary TIs. 

\subsection{Surface states in Thallium-based ternary compounds}

The thin films of Tl-based TIs are built from the bulk hexagonal structure with the atomic layers stacked along the crystallographic {\it z} direction and a net vacuum size of 3 nm above the top layer and below the bottom layer within a periodic simulation region. Recent theoretical\cite{lin} studies suggest contributions to the surface band structure from the dangling bonds arising from the surface terminations, and that these dangling states appear along with the topological surface states. Among four surface terminations, with the atomic layer ordering as Tl-Se(Te)-Bi-Se(Te), Se(Te)-Bi-Se(Te)-Tl, Bi-Se(Te)-Tl-Se(Te) and Se(Te)-Tl-Se(Te)-Bi, the one with Se(Te) beneath the Tl layer is argued to have minimal surface dangling bond densitys\cite{lin}. Therefore we chose this termination in all further thin film studies.  We note here that experiments do not observe any signature of dangling bond states in the optical spectra\cite{ando}. And, at least for this termination, we do not find any signature of the dangling bond state either, at least, within the Dirac cone region. Different thicknesses of the thin films of TlBiSe$_2$, ranging from 23 to 39 layers, corresponding to thicknesses of 4 to 7 nm and 23 to 31 layers for TlBiTe$_2$ with the maximum thickness value of 5.8 nm were considered. Figure 4(a) shows a representative supercell structure of 39 layers of TlBiSe$_2$. We consider the relaxation of atomic positions in the {\it z}-direction since it is argued that the positions of Se(Te) atoms are critical in determining the topological nature of the surface states\cite{lin}. The positions are assumed to be optimized when the {\it z}-component of the forces are smaller than the threshold value 0.015 eV/\AA.

As for the Bi-based binary compounds, Tl-based ternary compounds also show thickness dependent electronic structure. The Dirac cone is preserved in TlBiSe$_2$ with 39 atomic layers, corresponding to a thickness of 7 nm (Fig. 5(a)), and in TlBiTe$_2$  with 31 layers of a thickness 5.8 nm (Fig. 5(b)), where red circles denote surface state contributions and the solid black lines represent the bulk bands. We reproduce the interesting experimental features of TlBiSe$_2$\cite{ando} and TlBi$_2$Te\cite{chen}. The bulk VB (BVB), in TlBi$_2$Se has two maxima between $\bar{\Gamma}$ and \={M} points in the hexagonal BZ (Fig. 4(b)) and the bulk CB (BCB) has a minimum at \={M}. The Dirac cone is found to exist inside the bulk band gap, at the $\bar{\Gamma}$-point of the BZ. These features are clearly seen in our computed band structure (Fig. 5(a)). Our computed band structure of TlBiTe$_2$ shows that the Dirac point is below the BVBM (Fig. 5(b)). The BCBM is placed at $\bar{\Gamma}$ and the BVBM along the line joining $\bar{\Gamma}$ and \={M}, producing an indirect gap of 83 meV. Another BVBM emerges along the line joining $\bar{\Gamma}$ to \={K}. These features including the indirect gap size, positions of BVB, BCB and the Dirac point are consistent with the recent experimental study\cite{chen}.          

For thin film thicknesses corresponding to fewer than 39 (31) atomic layers in TlBiSe$_2$ (TlBiTe$_2$), our study suggests that a band gap opens in the otherwise metallic surface states, much like those in binary Bi-based TIs\cite{zhang2}. With decreasing thicknesses, the size of the gap seems to increase indicating the increased interactions between surfaces states that exist on two opposite surfaces of the TI (Figures not shown). The gap values are summarized in Table I. There are no experimental reports, so far, on the critical film thickness necessary to preserve the metallic bands for Tl-based ternary TIs which our theoretical study predicts. It is interesting to note that the predicted induced gaps in Tl-based TIs, for the same film thickness, are larger than the binary Bi-based TIs \cite{jiwon}.    

\begin{figure}
\scalebox{0.315}{\includegraphics[angle=-90]{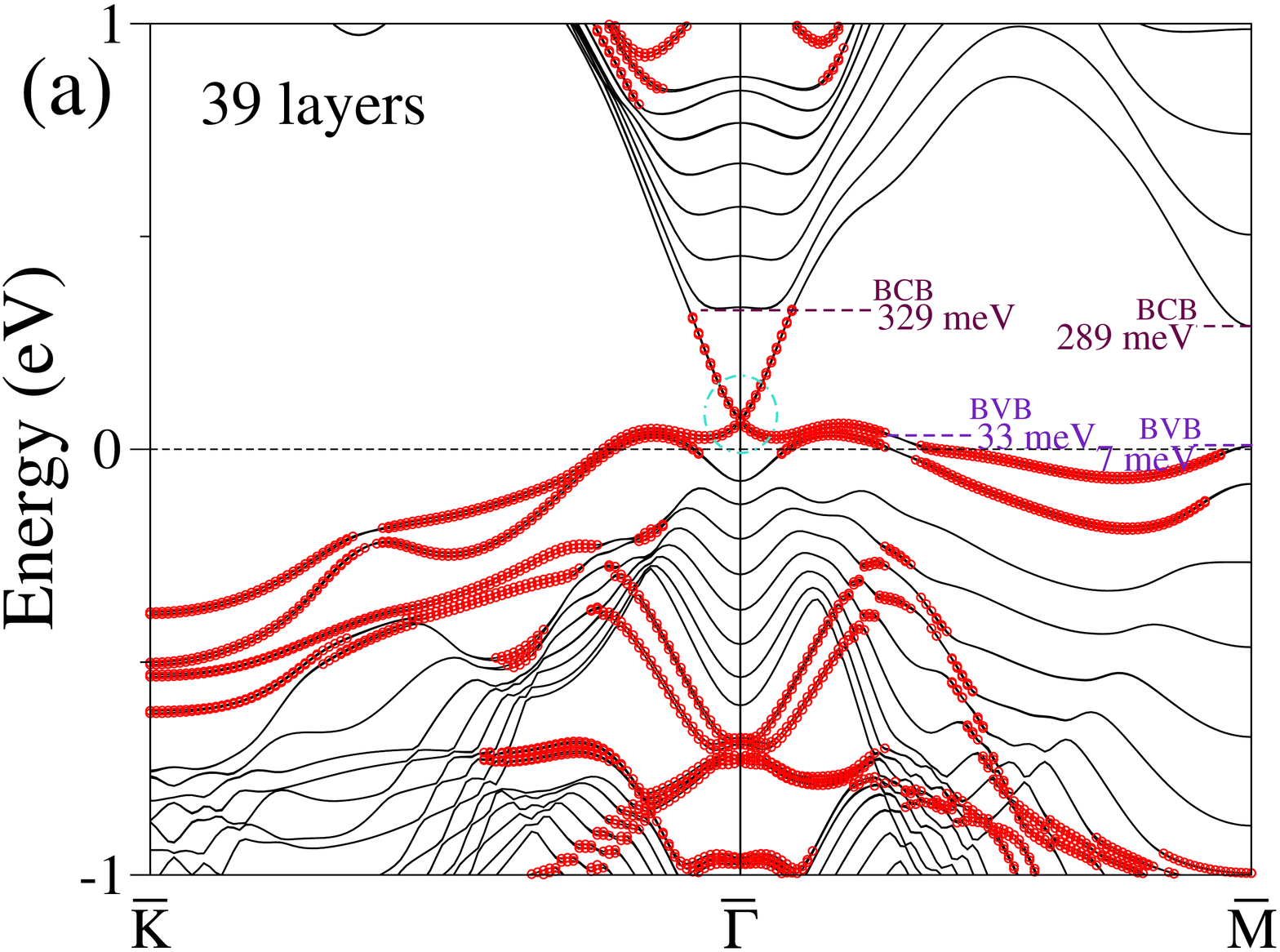}}
\scalebox{0.315}{\includegraphics[angle=-90]{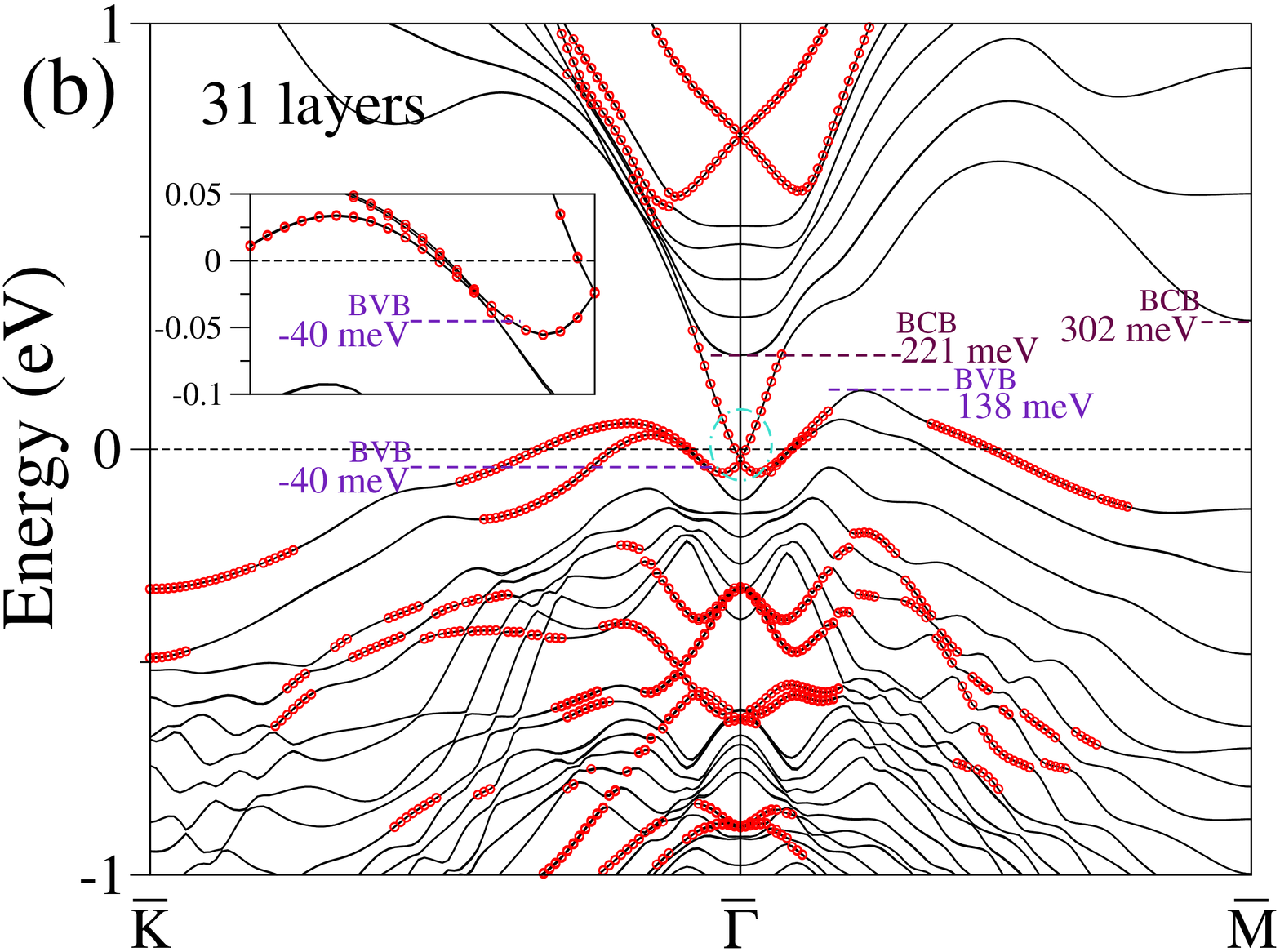}}
\caption{ (Color online) Band structures of (a) TlBiSe$_{2}$ film with a thickness of $\sim$7.0 nm (39 layers) and (b) TlBiTe$_{2}$ film with a different thickness of $\sim$5.8 nm (31 layers). Thinner films (23, 27, 31, 35 layers for TlBiSe$_{2}$ and 23, 27 layers for TlBiTe$_{2}$) show nonzero band gaps at $\bar{\Gamma}$ (Figures not shown). Orbital contributions from first few layers in each thicknesses are marked with red circles.}
\label{fig:Fig5}
\end{figure}

\begin{figure}
\scalebox{0.315}{\includegraphics[angle=-90]{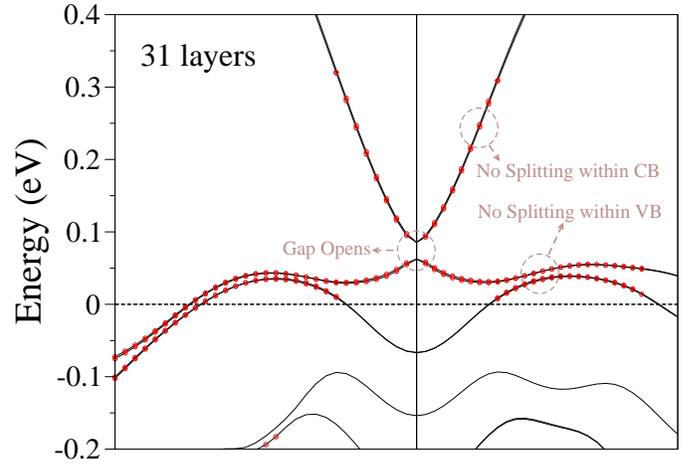}}
\caption{ (Color online) Band structure of TlBiSe$_{2}$ film with a thickness of $\sim$5.6 nm (31 layers). A nonzero band gap exists at $\bar{\Gamma}$ due to the coupling between the top surface CB and the bottom surface VB and between the top surface VB and the bottom surface CB. No splitting is observed away from the surface state band edges. 
 }
\label{fig:Fig6}
\end{figure}

\begin{figure}
\scalebox{0.345}{\includegraphics[angle=0]{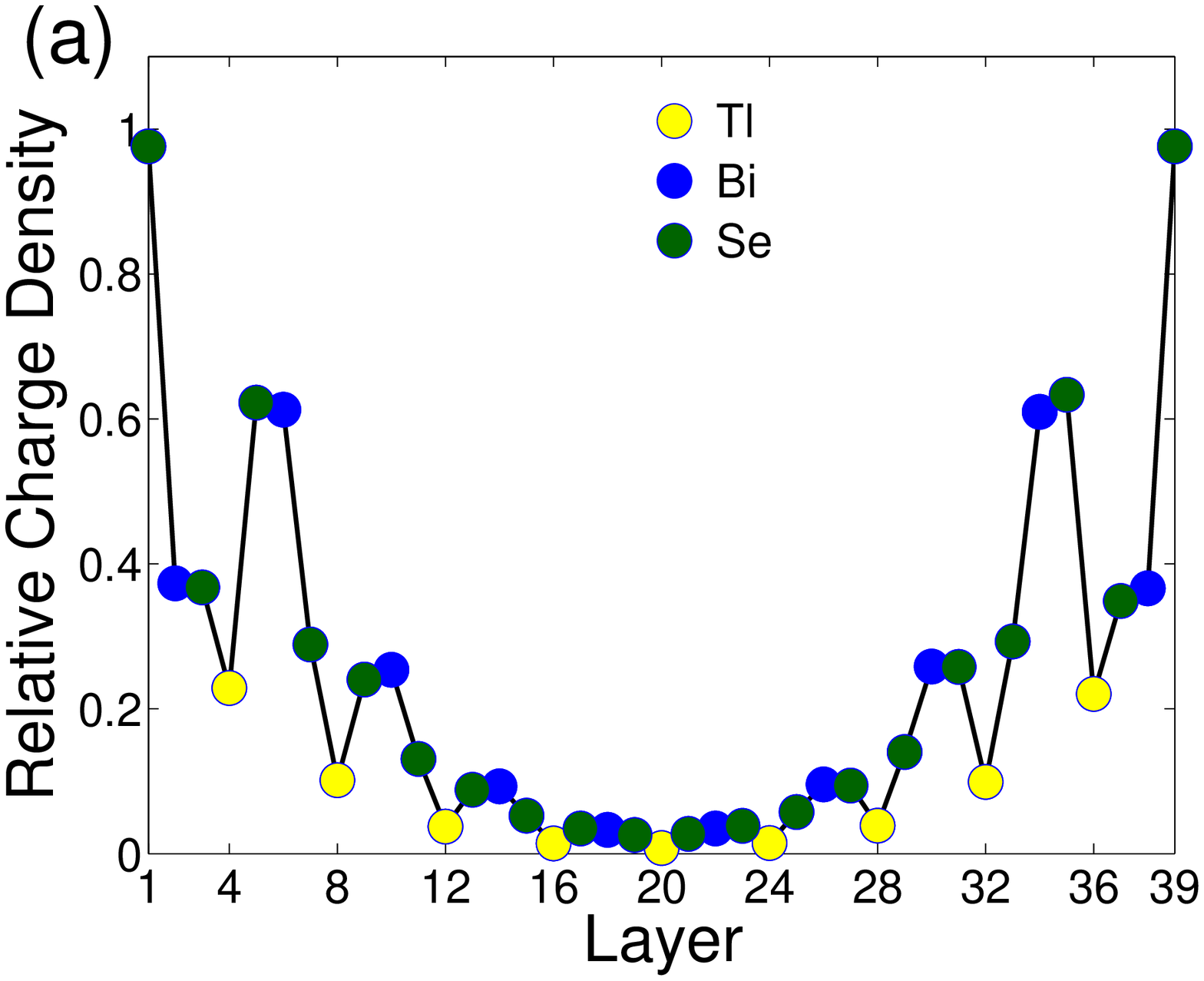}}
\scalebox{0.345}{\includegraphics[angle=0]{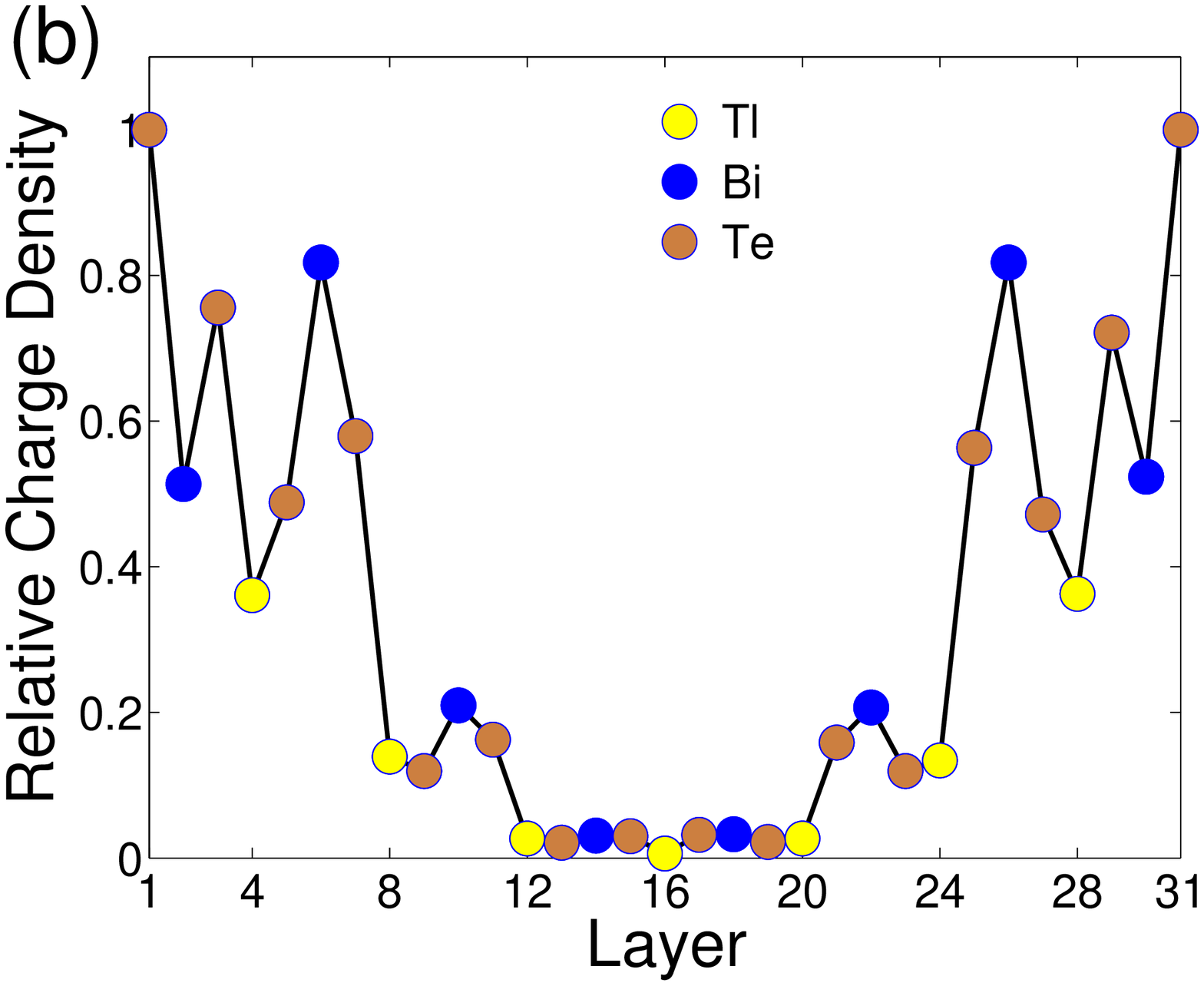}}
\caption{ (Color online) Layer-projected charge density, contributed by surface states in the neighborhood of the $\bar{\Gamma}$-point for (a) TlBiSe$_{2}$ with a thickness of $\sim$ 7 nm, corresponding to 39 atomic layers and (b) TlBiTe$_2$ with a thickness of $\sim$ 5.8 nm, corresponding to 31 layers. The color labeling scheme is consistent with that in Figure 1 (a).}
\label{fig:Fig7}
\end{figure}

\begin{table}
\caption{The induced band gap at the time-reversal invariant point $\bar{\Gamma}$  $\Delta$E$_{\bar{\Gamma}}$ (in eV) for thin-films with various thicknesses.}
\begin{tabular}{ c | c | c | c | c | c  }
\hline \hline
  Number of Layers & 23 & 27 & 31 & 35 & 39 \\
\hline
  $\Delta$E$_{\bar{\Gamma}}$  & 0.059 & 0.042 & 0.023 & 0.015 & 0.000\\
  TlBiSe$_{2}$ & & & & &\\
\hline
  $\Delta$E$_{\bar{\Gamma}}$  & 0.018 & 0.009 & 0.000 & NA & NA\\
 TlBiTe$_{2}$ & & & & &\\
\hline \hline
\end{tabular}
\end{table}

\begin{figure}
\scalebox{0.355}{\includegraphics[angle=0]{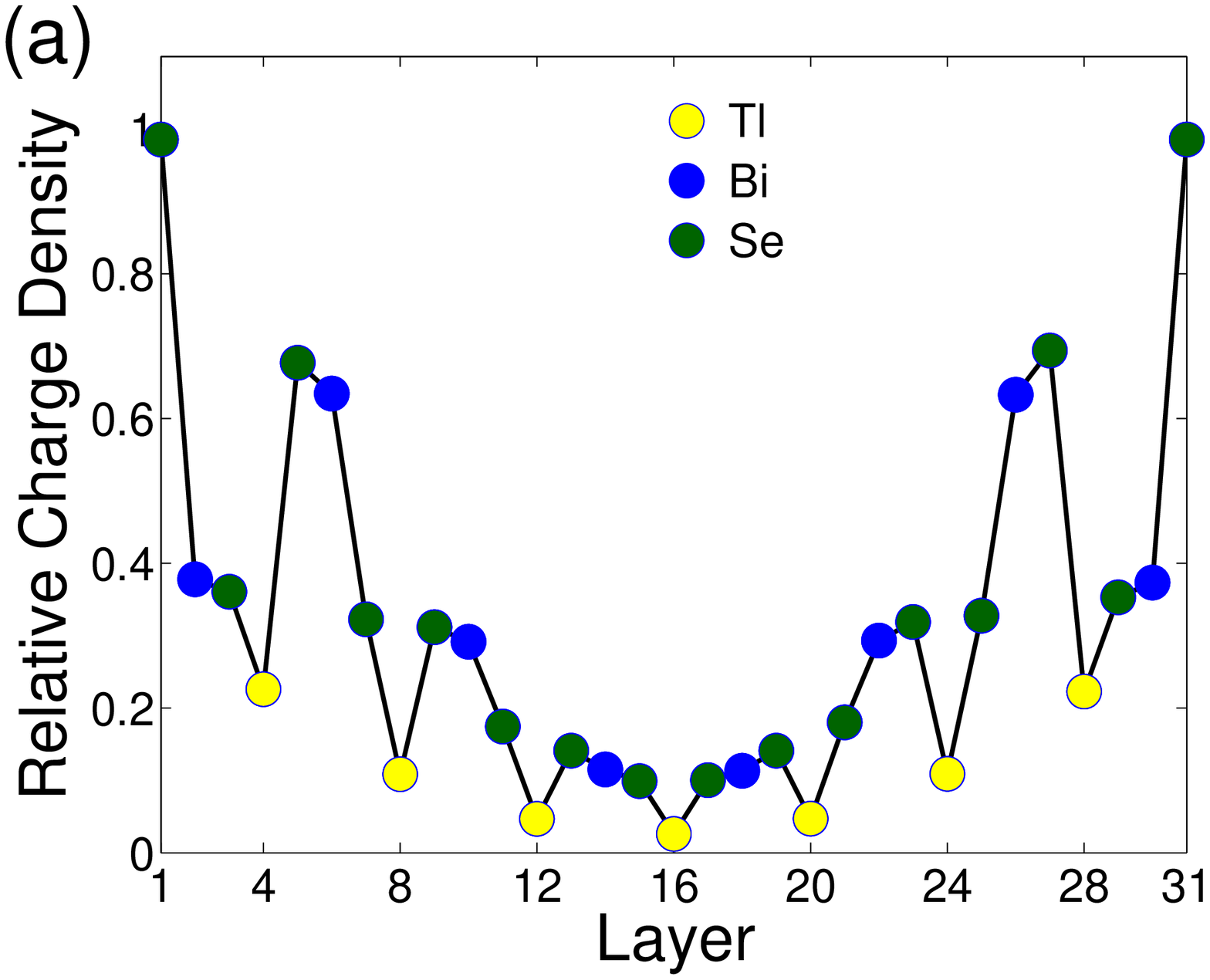}}
\scalebox{0.355}{\includegraphics[angle=0]{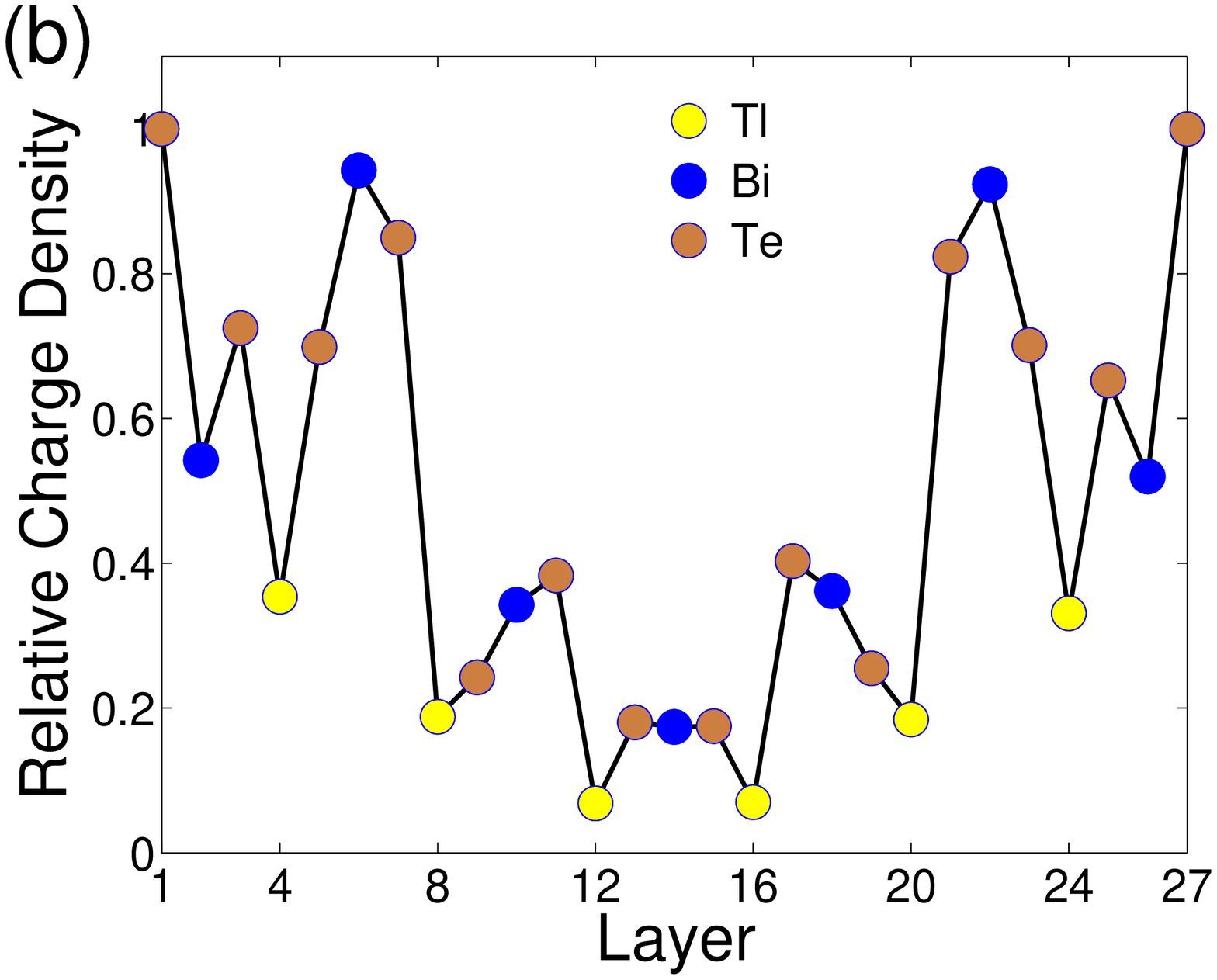}}
\caption{ (Color online) Layer-projected relative charge density for thinner samples of (a) TlBiSe$_{2}$ with thickness of $\sim$ 5.6 nm corresponding to 31 layers and (b) TlBiTe$_2$ with thickness of $\sim$ 5 nm corresponding to 27 layers. The color labelling scheme is consistent with that in Figure 1(a).}
\label{fig:Fig8}
\end{figure}

\begin{figure}
\scalebox{0.455}{\includegraphics[angle=0]{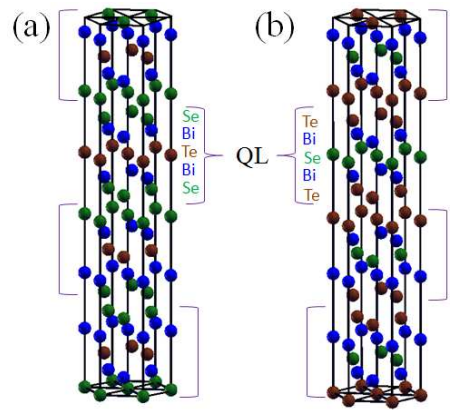}}
\caption{ (Color online) Schematic diagram of (a) Bi$_{2}$Se$_{2}$Te and (b) Bi$_{2}$Te$_{2}$Se thin film structures obtained by stacking 4QLs along {\it z}-direction.}
\label{fig:Fig9}
\end{figure}

To understand the origin of the gap at the Dirac point with decreasing film thickness, we first map out the surface state contributions in the band structure from atoms within first few layers of the film, and from all the orbitals, relative to the total contributions of all orbitals from all the layers of the thin film. From the top or bottom surface of the film, 8(6) atomic layers were chosen for TlBiSe$_2$(TlBiTe$_2$). A larger number of atomic layers are used for TlBiSe$_2$ than for TlBiTe$_2$ because a larger number of atomic layers is required to preserve the metallic band structure because the surface states are less localized to the surface region in TlBiSe$_2$. These choices also helped us to choose the same cut-off percentage for both the Tl-based compounds in estimating the surface state contributions. For film thicknesses corresponding to 23, 27, 31, 35 and 39 atomic layers in TlBiSe$_2$, respectively, these cut-offs were varied from 70$\%$ - 50$\%$, in steps of 5$\%$ decrease. For TlBiTe$_2$, 23, 27 and 31 atomic layers were considered with cut-offs of 70$\%$, 65$\%$ and 60$\%$. Examples results of these orbital contributions are marked on the top of the total band structure in Fig. 5(a) and (b).

After estimating the distribution of surface states in the thin film structure, we computed the valence charge density resulting from these surface state wave-functions in both the compounds. We focused on surface states around the Dirac-point, in an energy window chosen such that same number of wave-functions are used for building the surface charge density for most relevant thicknesses considered in this study.  The wave-functions at three closely spaced {\bf k}-points including the $\Gamma$-point were chosen. The charge density is computed layer-wise. We define the {\it relative} charge density, to be the ratio of peak charge density in a particular layer to the peak charge density in all layers. This quantity is plotted in Figs. 7(a) and (b). Our studies suggest exponential decay of the charge density in the bulk region but with slower decay in TlBiSe$_2$ than in TlBiTe$_2$. In about 12-14 atomic layers, from top or bottom surfaces of the film (corresponding to 2.3 nm), the surface state contributions decay to almost zero in TlBiTe$_2$, whereas, the surface states extends up to 16-18 atomic layers ($\sim$3 nm) in TlBiSe$_2$. With decreasing film thickness, the charge density accumulates in the middle of bulk region in both compounds (Figs. 8(a) and (b)) suggesting increasing interactions of surface states by overlap of their wave-functions inducing a gap at the Dirac point. However, while a band gap is opened in the surface state band structure, away from the surface state band edges band structure is largely unaffected as seen in Fig. 6 -i.e., it does not split into two bands as one would expect from resonant coupling between opposite surface states-suggesting that the splitting is due to interband coupling between the top surface CB and the bottom surface VB and between the top surface VB and the bottom surface CB.      

\begin{figure}
\scalebox{0.315}{\includegraphics[angle=-90]{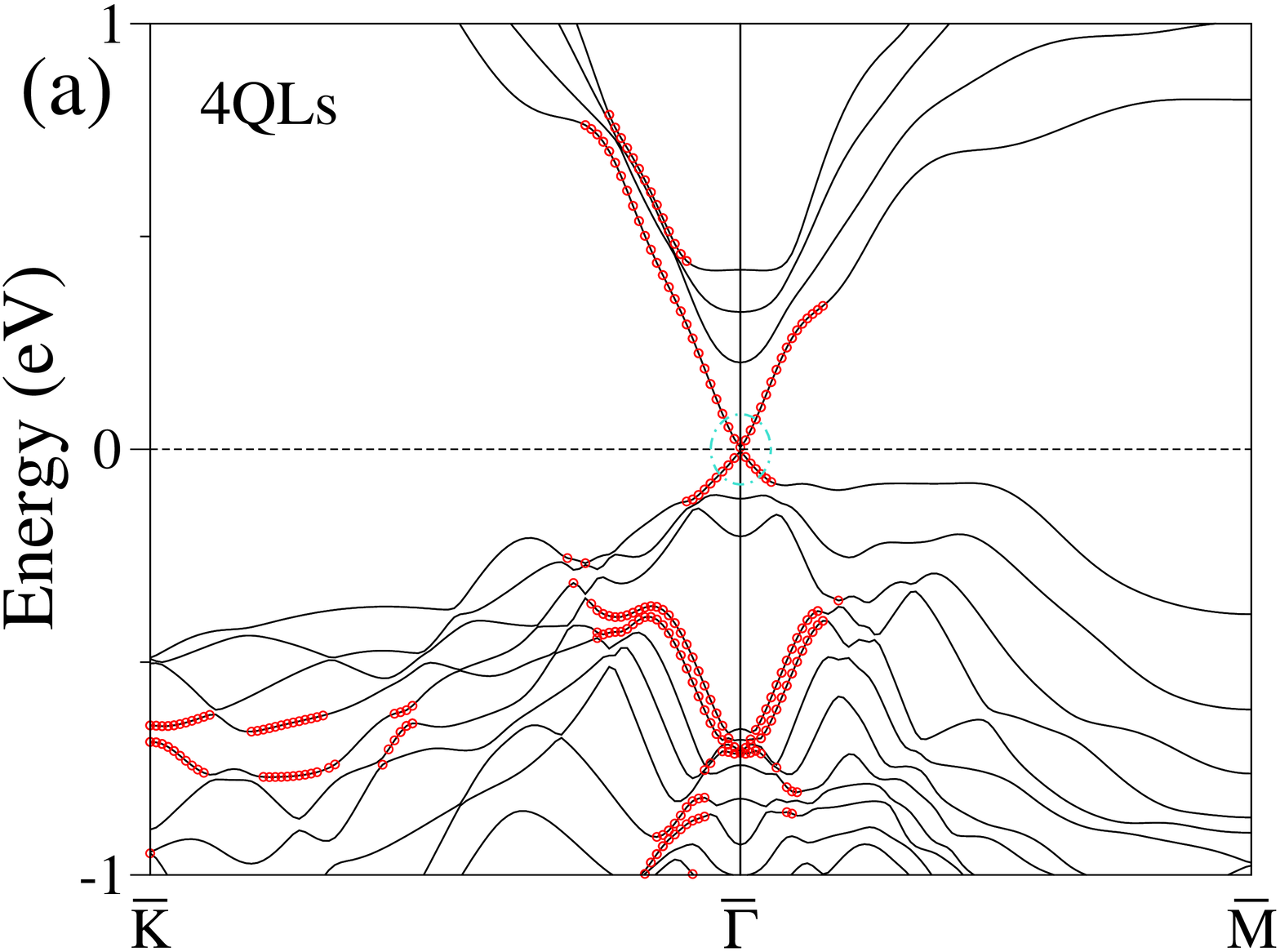}}
\scalebox{0.315}{\includegraphics[angle=-90]{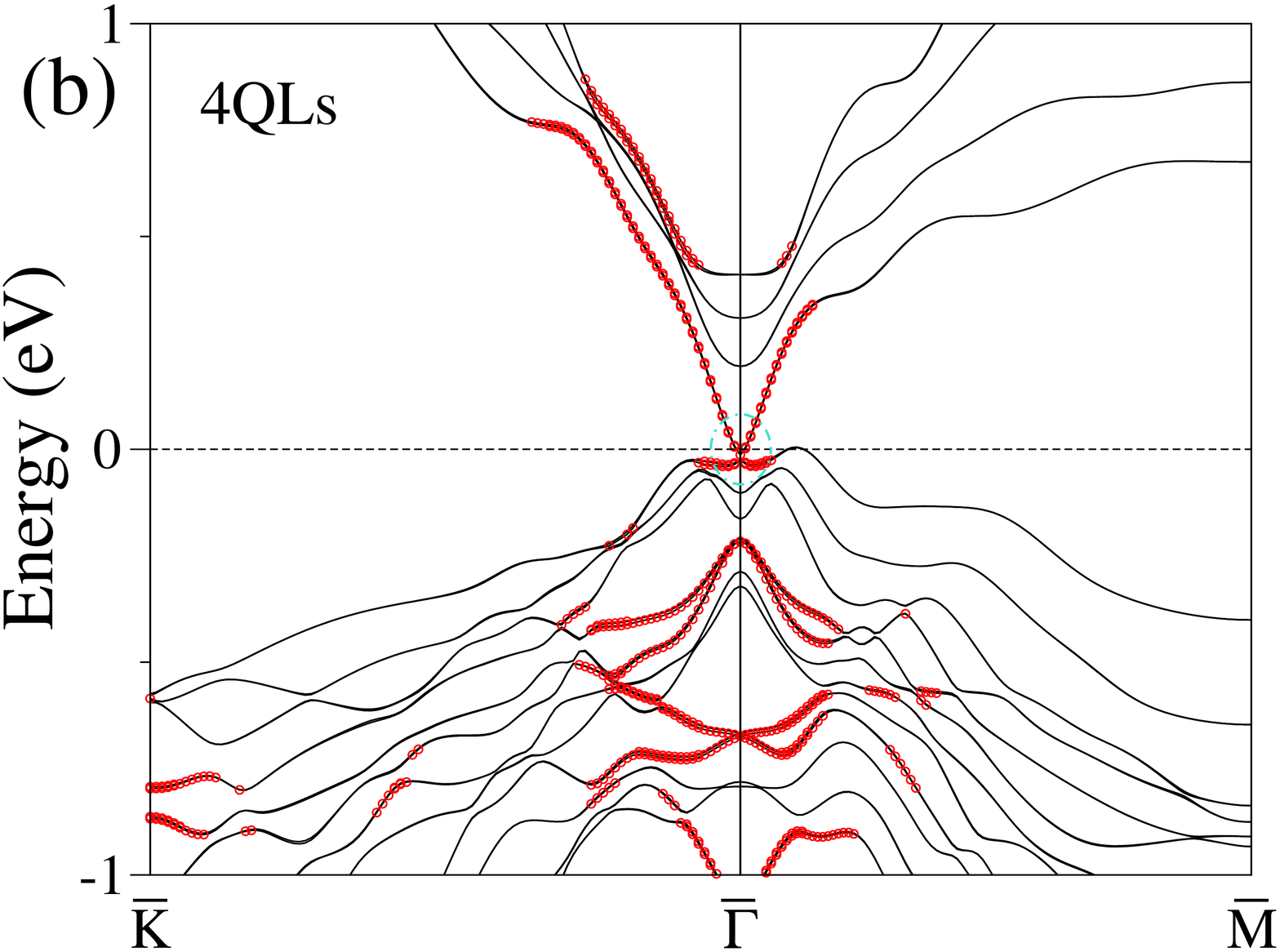}}
\caption{ (Color online) Band structures of Bi$_{2}$Se$_{2}$Te films with thickness of 3.6 nm corresponding to 4 QLs (a) without atomic relaxation and (b) with atomic relaxation. The Dirac cone is affected by relaxing the atoms.}
\label{fig:Fig10}
\end{figure}

\begin{figure}
\scalebox{0.315}{\includegraphics[angle=-90]{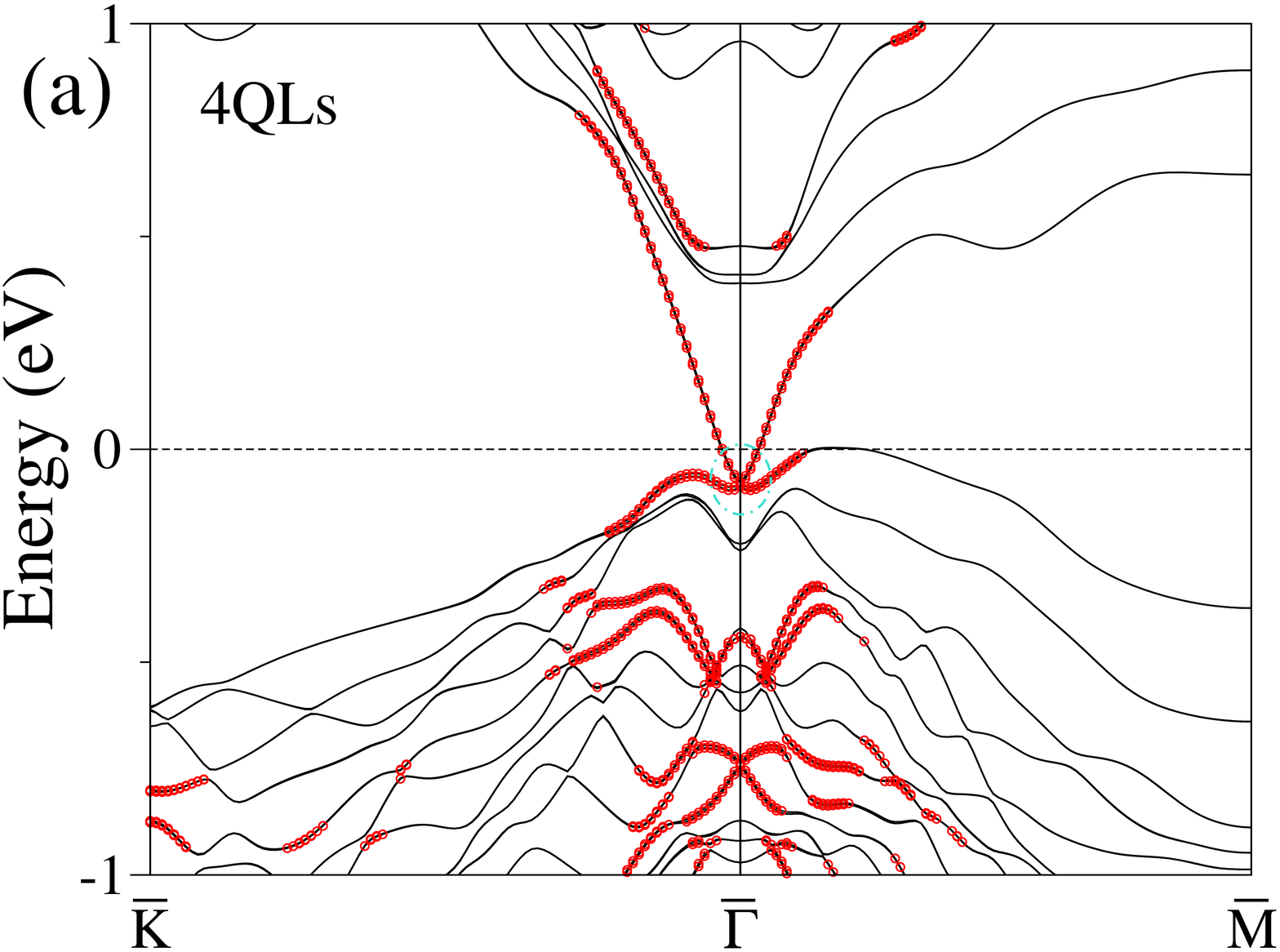}}
\scalebox{0.315}{\includegraphics[angle=-90]{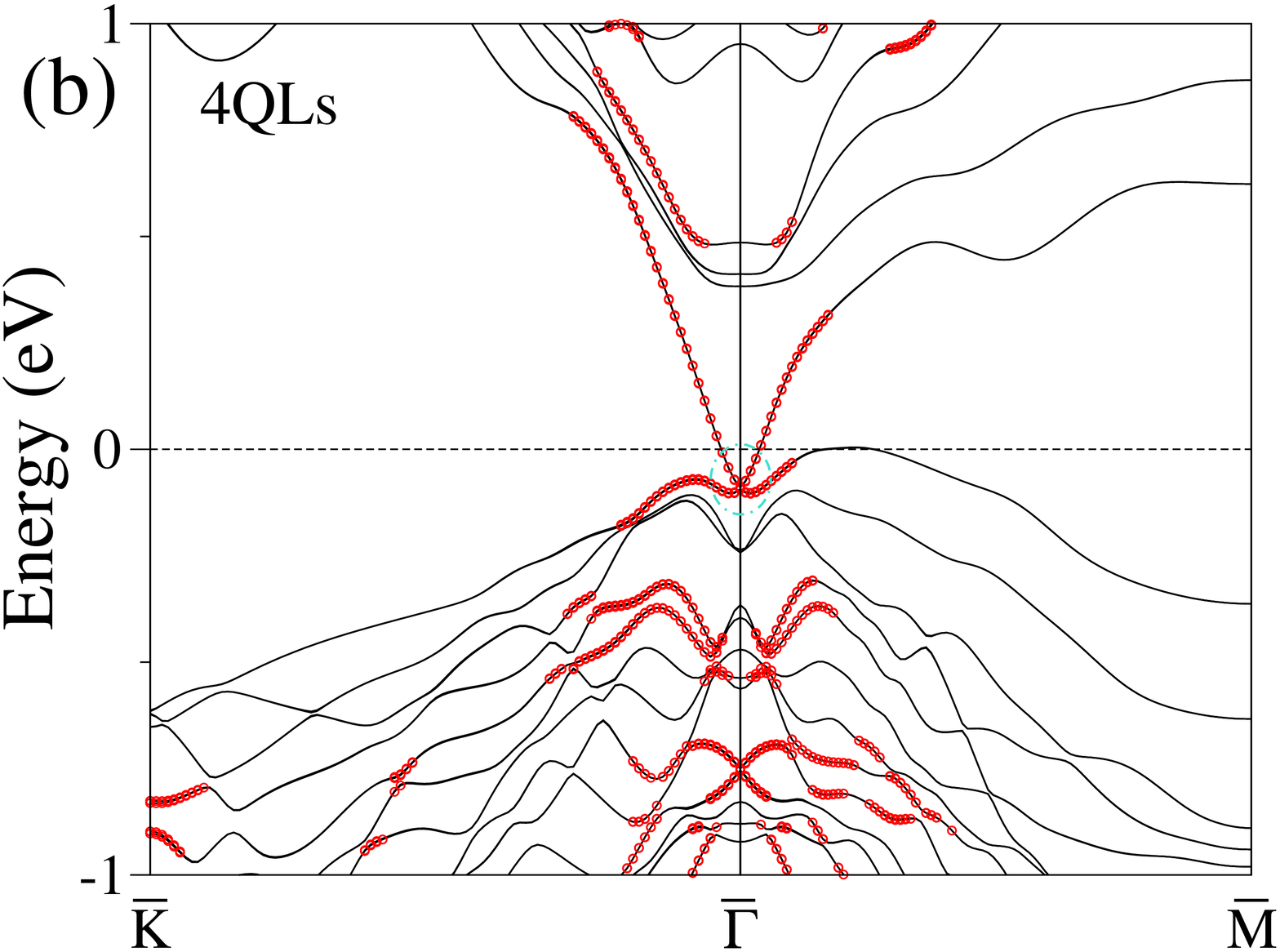}}
\caption{ (Color online) Band structures of Bi$_{2}$Te$_{2}$Se films with thickness of $\sim$ 3.7 nm corrresponding to 4 QLs (a) without atomic relaxation and (b) with atomic relaxation. The Dirac cone is not affected by relaxation.}
\label{fig:Fig11}
\end{figure}

\begin{figure}
\scalebox{0.345}{\includegraphics{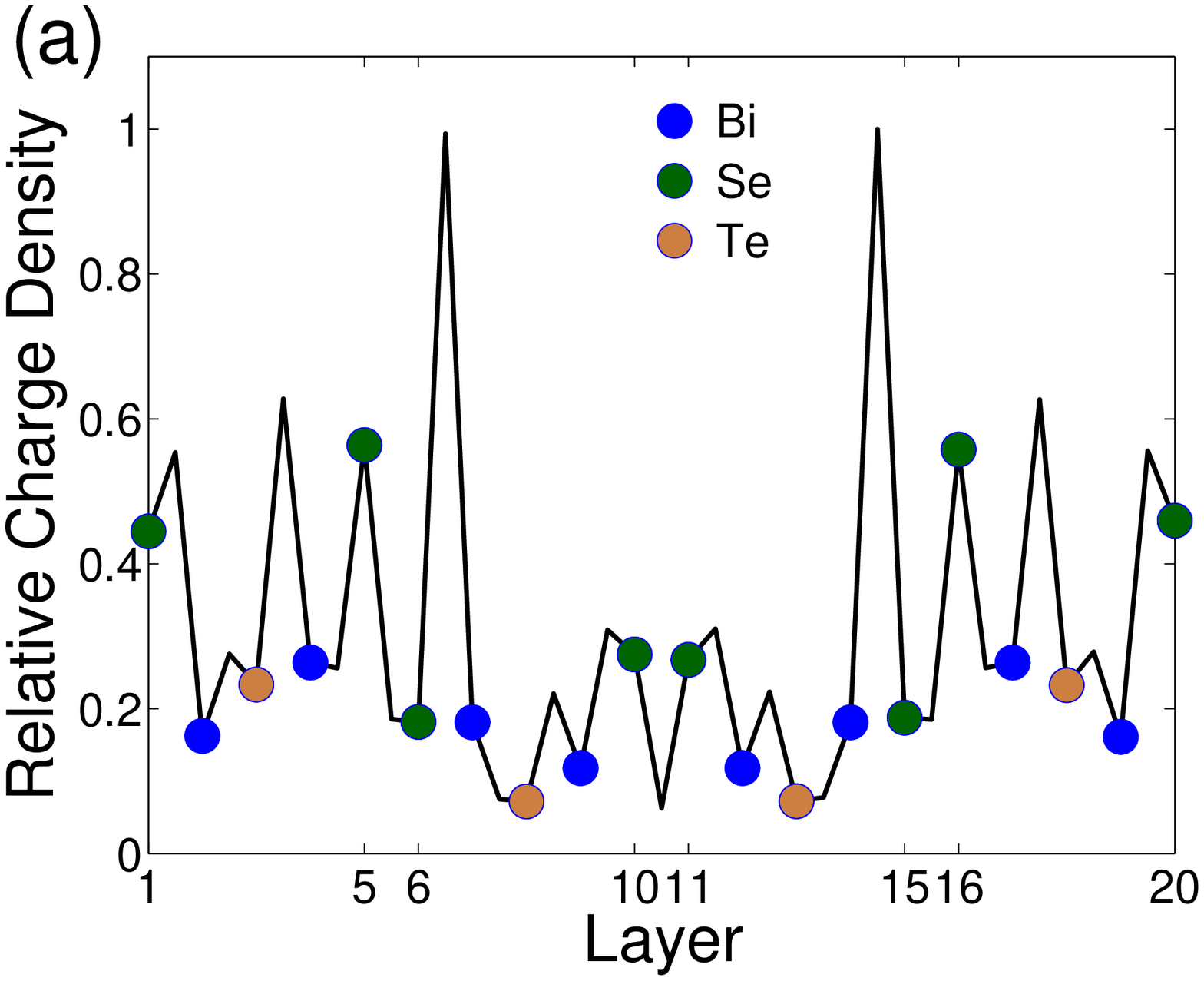}}
\scalebox{0.345}{\includegraphics{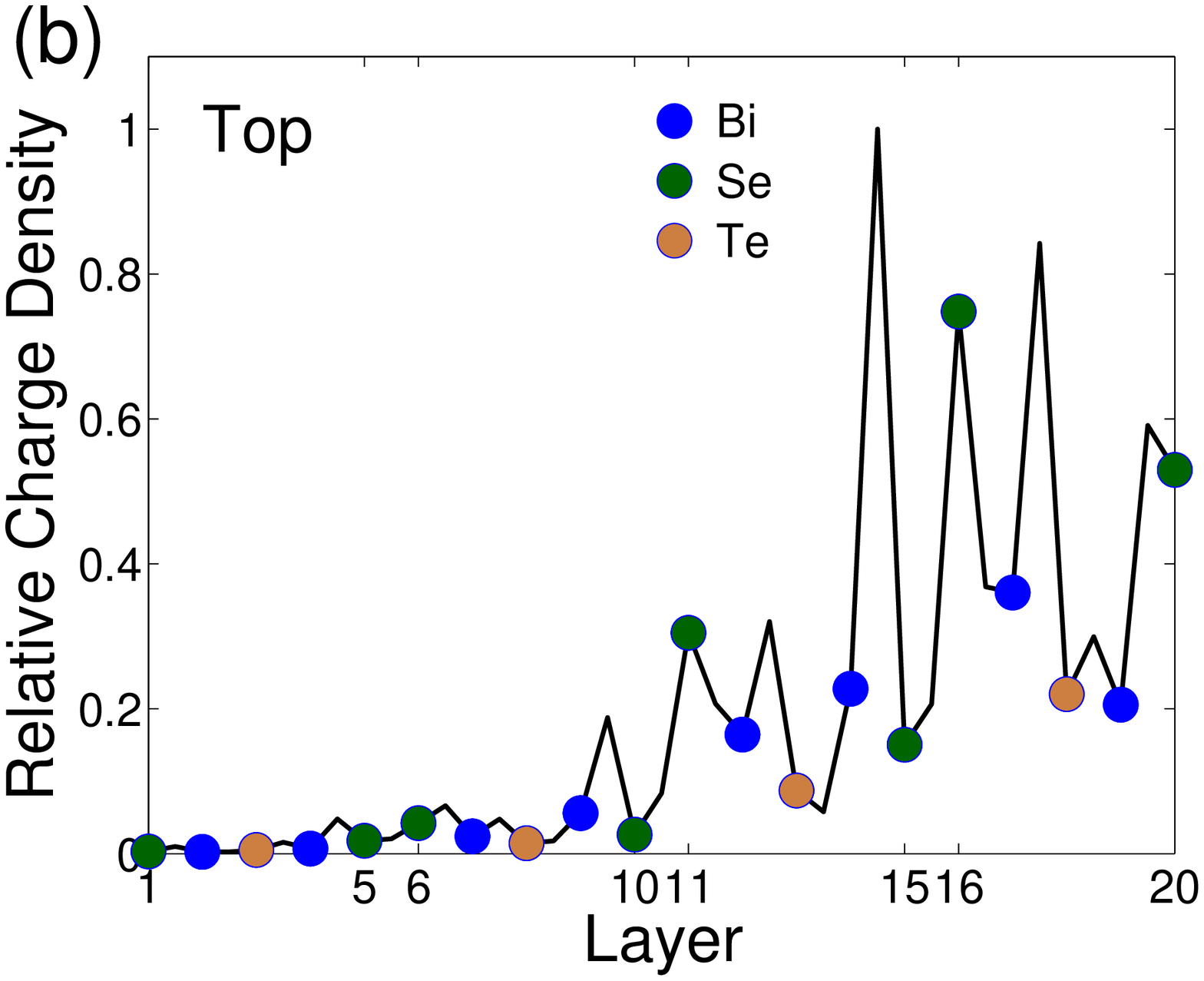}}
\scalebox{0.345}{\includegraphics{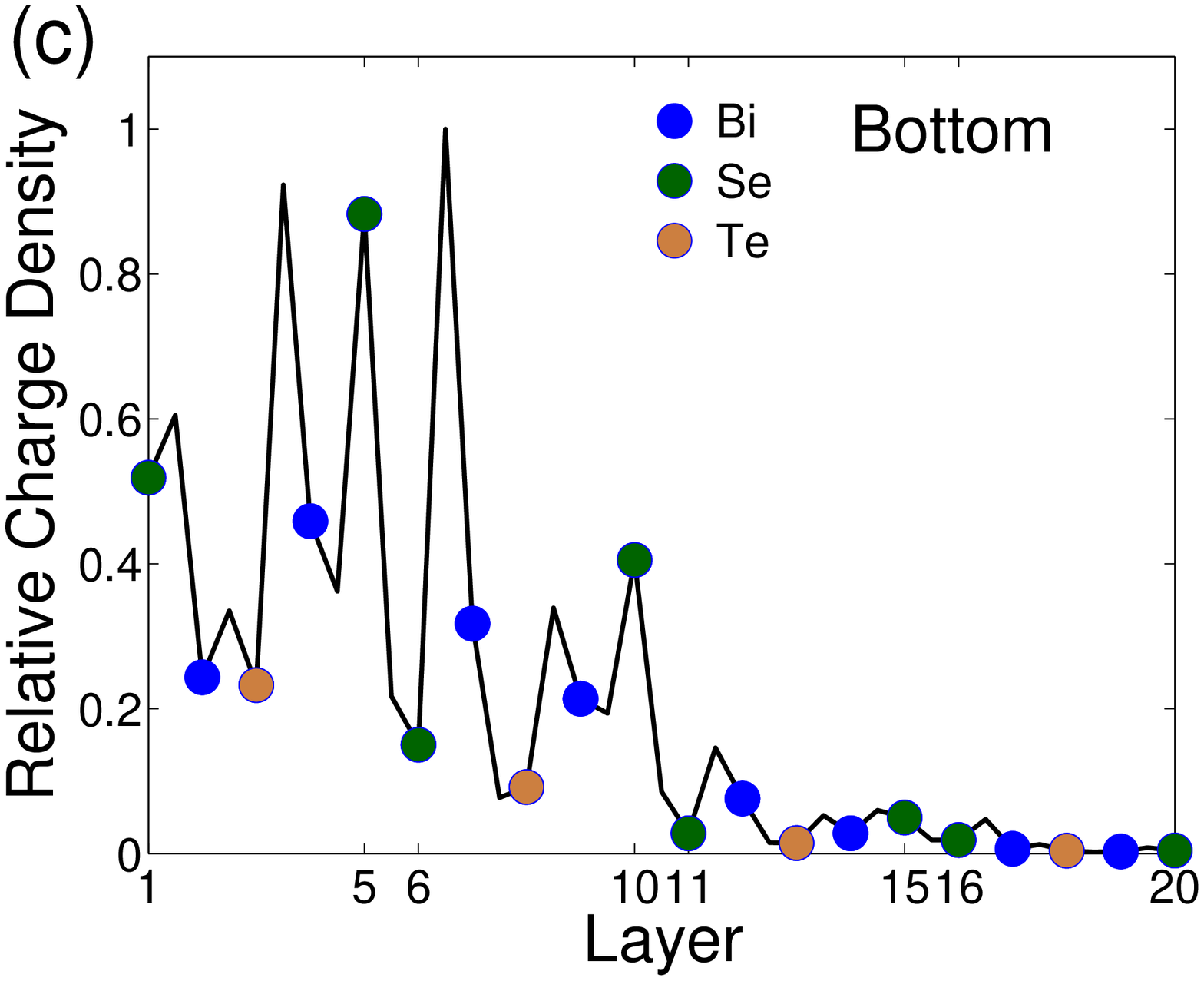}}\\
\scalebox{0.345}{\includegraphics{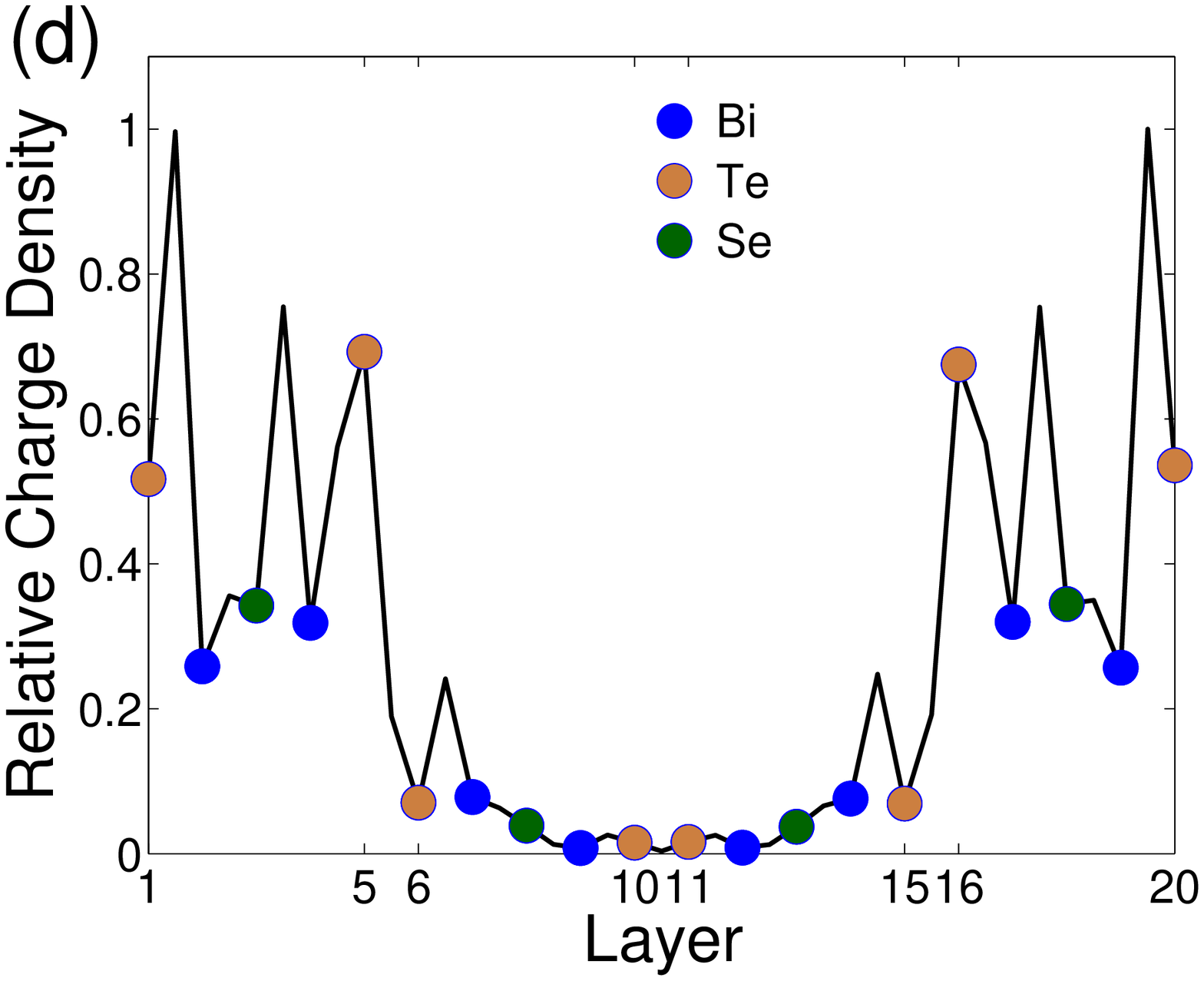}}\\
\caption{ (Color online) (a) Layer-projected charge density for Bi$_{2}$Se$_{2}$Te without the atomic relaxation, contributed by surface states in the neighborhood of the $\bar{\Gamma}$-point for thickness of $\sim$3.6 nm corresponding to 4 QLs. Charge density resulting from (b) top and (c) bottom surface wave functions and (d) Bi$_{2}$Te$_{2}$Se with the atomic relaxation for the thickness of $\sim$3.7 nm corresponding to 4 QLs. Although a charge pile up in the middle is seen in (a), the separate wavefunction contributions in (b) and (c) clearly show that surface charge density strictly vanishes after 2 QLs from both the top and bottom surfaces of the film.}
\label{fig:Fig12}
\end{figure}

\subsection{Surface states of Bismuth-based ternary compounds} 

We discuss the surface state properties of Bi$_2$Se$_2$Te and Bi$_2$Te$_2$Se in this section. The building blocks of these compounds are QLs, arranged in the order -Se(Te)-Bi-Te(Se)-Bi-Se(Te)-, and stacked along the crystallographic {\it z}-direction (Figs. 9(a) and (b)). Different film thicknesses corresponding the different number of QLs were considered with a maximum of 4 QLs. The choice thickness is guided by the necessity to maintain the metallic nature of the surface bands. The computational parameters used in this study are same as those for Tl-based TIs. We first construct the thin film from experimental parameters\cite{nakajima} and then study the thin-film properties without and with relaxing the atomic positions with the force convergence criteria of 0.05 eV/\AA. The procedure for extracting surface states from crystal wave-functions and the construction of layer-wise surface charge density is same as that for Tl-based TIs discussed in the previous section.     

The thin-film structure of Bi$_2$Se$_2$Te, without internal position relaxation, shows a more symmetric Dirac cone with metallic bands, as compared to the Tl-based TIs for thickness corresponding to 4 QLs (Fig. 10 (a)). The bulk band gap (of 313 meV) supports these novel states, and the Dirac point is well within this gap. Recently, a report of 3 QLs, as the minimum thickness, required to preserve the Dirac cone has appeared\cite{johnson}. We find 4 QLs to be necessary, which we also illustrate with the help of surface charge density calculations. However, in the absence of any experimental work, it is likely that the minimum thickness, carefully predicted by different theoretical methods, may fall in the range of 3QLs-4QLs. For film thicknesses below 4 QLs, a finite gap is induced at the Dirac point whose size increases with decreasing the film thickness (Figures not shown). The size of the gap values are listed in Table II. With atomic relaxation, the Dirac cone is affected and its symmetry is lost (Fig. 10(b)). The BVB lies along the line joining $\bar{\Gamma}$ and \={M} in the hexagonal BZ, and its maximum is positioned almost in line with the Dirac point. This hints at crucial role played by rearrangement of atomic positions either {\it insitu} in thin-film preparation procedure or in presence of external environments. In Bi$_2$Te$_2$Se, a significant part of surface state band structure  falls below the BVBM and the Dirac point occurs in the energy neighborhood of the $\Gamma$-point. Atomic relaxation has almost negligible effect on the overall band structure (Figs. 11 (a) and (b)) in this case. Decreasing film thickness induces a finite gap (Table II). These studies suggest that the Dirac cone remains protected in thinner layers of ternary Bi-based TIs than in binary Bi-bsed TIs or than in Tl-based TIs.
   
The origin of the critical thickness needed to maintain the metallic nature of the surface states can be understood by studying the layer-dependent surface charge densities. The procedure is same as that used for Tl-based TIs, except the charge density in the interstitial regions is also considered in order to locate the surface state spread in the bulk region as precisely as possible for thinner films thicknesses. For 4 QL thick Bi$_2$Se$_2$Te films, the combined charge density from the surface states of  both sides of the  film seems to be  significant in the middle of the bulk region (Fig. 12(a)).  However, we separately computed the charge  densities resulting from top and  the bottom surface  states.  As  seen  from  the plots (Figs. 12(b) and (c)), the density associated with surface states drops rapidly beyond  2 QLs, indicating negligible  interaction  between  these opposite surface states  in the 4QL film, consistent with the observed protected Dirac cone. Relaxation of atomic positions is found  to have  no significant effect.   In the 4QL films of Bi2 Te2 Se, even the combined contribution from both surfaces is small in our  studies with and  without atomic relaxation (Figs. 12(d)).

\begin{table}
\caption{The induced band gap at the time-reversal invariant point $\bar{\Gamma}$  $\Delta$E$_{\bar{\Gamma}}$ (in eV) for thin-films with various thicknesses.}
\begin{tabular}{ c | c | c | c }
\hline \hline
  Number of QLs & 2 & 3 & 4 \\
\hline
  $\Delta$E$_{\bar{\Gamma}}$  & 0.128 & 0.038 & 0.000 \\
  Bi$_{2}$Se$_{2}$Te without relaxation & & & \\
\hline
  $\Delta$E$_{\bar{\Gamma}}$  & 0.086 & 0.017 & 0.000 \\
  Bi$_{2}$Se$_{2}$Te with relaxation & & & \\
\hline
  $\Delta$E$_{\bar{\Gamma}}$  & 0.039 & 0.037 & 0.000 \\
  Bi$_{2}$Te$_{2}$Se without relaxation & & & \\
\hline
  $\Delta$E$_{\bar{\Gamma}}$  & 0.027 & 0.037 & 0.000 \\
  Bi$_{2}$Te$_{2}$Se with relaxation & & & \\
\hline \hline
\end{tabular}
\end{table}

\section {Summary and Conclusions}
We use a density functional based electronic structure method to study the thin film surface state properties of Tl and Bi-based ternary topologicial insulators. These studies predict that the Dirac cone remains protected in thinner layers of ternary Bi-based TIs than in binary Bi-based TIs or than in Tl-based TIs, which may be advantageous for certain applications. However, we also predict that in the atomically relaxed strucures, the Dirac cone of all but Bi$_2$Se$_3$ lie in or near the BVB. However, the Dirac cone of Bi$_2$Se$_2$Te remains outside of the BVB in the absence of relaxation, as might occur with dielectrics replacing the air gap. And, of course, the bonding to the dielectric could directly affect the surface states as well. Our computed results agree very well with experimental results where available, while other predictions such as the critical film thicknesses required to maintain the Dirac cone, size of the induced gaps, and the extent of the spread of surface states in the bulk region point to further needed experimental work.

\acknowledgments
The authors acknowledge financial support from the Nanoelectronics Research Initiative supported Southwest Academy of Nanoelectronics (NRI-SWAN) center. We thank the Texas advanced computing center (TACC) for computational support (TG-DMR080016N).

\end{document}